\documentclass[preprint,superscriptaddress,showpacs,nofootinbib,onecolumn]{revtex4-1}

\pdfoutput=1

\usepackage{hyperref}
\usepackage{graphicx}
\usepackage{longtable}
\usepackage{revsymb}

\begin{document}

\title{\bf The dependence of the atomic energy levels on a superstrong
  magnetic field with account of a finite nucleus radius and mass}

\author{S~.I.~Godunov}
\email{sgodunov@itep.ru}
\affiliation{\small Institute for Theoretical and Experimental Physics,
  117218, Moscow, Russia}
\affiliation{\small Novosibirsk State University, 630090, Novosibirsk, Russia}
\author{M.~I.~Vysotsky}
\email{vysotsky@itep.ru}
\affiliation{\small Institute for Theoretical and Experimental Physics,
  117218, Moscow, Russia}
\affiliation{\small Novosibirsk State University, 630090, Novosibirsk, Russia}
\affiliation{\small Moscow Engineering Physics Institute, 115409, Moscow, Russia}
\affiliation{\small Moscow Institute of Physics and Technology, 141700, Dolgoprudny, Moscow 
Region, Russia}

\begin{abstract}
  The influence of the finiteness of the proton radius and mass on the
  energies of a hydrogen atom and hydrogen-like ions in a superstrong
  magnetic field is studied. The finiteness of the nucleus size pushes
  the ground energy level up leading to a nontrivial dependence of the
  value of critical nucleus charge on the external magnetic field.
\end{abstract}


\maketitle

\section{Introduction}

At magnetic fields $B\geq (3\pi/e^{2})B_{0}\equiv 3\pi
m_{e}^{2}/e^{3}\approx 6\cdot 10^{16}~ \rm G$\footnote{We use the
  system of units in which $\hbar=c=1$, $\alpha=e^{2}=1/137.03599\dots$} the Coulomb potential of the
nucleus becomes screened due to large radiative corrections \cite{1} ($B_{0}\equiv
m_{e}^{2}/e=4.4\cdot 10^{13}~ \rm G=4.4\cdot 10^{9}~ \rm T$). This leads
in particular to the freezing of the ground state energy of the
hydrogen atom at the value $E_{0}=-1.7~ \rm keV$ \cite{1,2}.
This statement is correct up to the values of the magnetic field at which
the Landau radius $a_{H}\equiv 1/\sqrt{eB}$ becomes close to the radius of
the nucleus, $a_{H}\approx R$. For hydrogen this happens at
$B\approx 10^{5}B_{0}\approx 10^{19}~\rm G$, where the value of the proton charge radius
$R=0.877~\rm fm$ (see \cite{pdg}) was used for numerical estimate. The approximation
of a pointlike proton is not valid for $B\gtrsim 10^{19}~\rm G$, and
in Sec. \ref{sec:potential} and \ref{sec:finsize} we will find to what
changes of atomic energies the finiteness of the proton size leads.

With the growth of the nuclei charge $Z$ the energy of the ground
electron level diminishes and in the absence of magnetic field a nucleus
with $Z=172$ is critical: the ground level sinks into the lower
continuum. As soon as the charge of the ion reaches $Z_{cr}=172$ two
$e^{+}e^{-}$-pairs are produced from the vacuum. Electrons with opposite
spins occupy the ground energy level while positrons are emitted to
infinity \cite{3}. In an external magnetic field the value of $Z_{cr}$
diminishes \cite{4,5}. According to \cite{4} at $B\approx 10^{2}B_{0}$
uranium becomes critical, $Z_{cr}=92$, while at $B=10^{4}B_{0}$ even
comparatively light nuclei are critical, $Z_{cr}\approx 40$. These
results were obtained without taking the screening of the Coulomb
potential into account. It was accounted for in \cite{6} where 
it was found that because of screening larger magnetic fields are
needed for a particular nucleus to become critical. Even more:
according to \cite{6} nuclei with $Z<50$ do not reach criticality
because of screening. The approximation of pointlike nuclei (which is
valid when the Landau radius is considerably larger than the size of
the nucleus) was used in \cite{6}. A nucleus with $Z\approx 56$
becomes critical when the magnetic field is so large that the Landau
radius equals the size of the nucleus. It was noted in \cite{6} that
the diminishing of the Coulomb potential due to the finite radius of
the nucleus should push the value of the ground energy level up, and
it was expected that this phenomenon will prevent ions with $Z=50,51$
from reaching criticality. In Sec. \ref{sec:finsizeZ} we present
quantitative consideration on the influence of the finiteness of
nucleus size on the dependence of $Z_{cr}$ on the value of the
external magnetic field $B$. This dependence turns out to be not so
simple. In particular, only nuclei with
$Z>59$ reach criticality in a superstrong magnetic fields. When $B$
further grows even nuclei with $Z>59$ become noncritical (see
Fig. \ref{fig:levels}). 

In \cite{1,2} and \cite{6} the atomic nucleus was considered as an 
infinitely heavy source of the Coulomb field. Because of the 
finiteness of the mass of the nucleus its motion in the magnetic field
should be taken into account and one should consider the two body
(electron and nucleus) problem in the presence of a homogeneous
magnetic field. This consideration was made in
\cite{7,7a,8}. According to the results obtained in \cite{8} hydrogen  
atomic levels get increased by $e|m|B/m_{p}$, where $m$ is
the projection of the relative angular momentum on the direction of
the external magnetic field and $m_{p}$ is the proton mass. The
corresponding formulae and numerical estimates are presented in 
Sec. \ref{sec:finmass}. 

The finiteness of the nucleus mass also leads to a nonzero hyperfine
interaction between the spins of the proton and the electron. Its
importance in the case of a superstrong $B$ was stressed in \cite{9}
and we present our comments in Sec. \ref{sec:hyperfine}.

\section{Electric potential of the nucleus }
\label{sec:potential}

The finite size of the nucleus makes the Coulomb potential less singular
at small distances, pushing up the electron levels. However the shape of
potential at distances much smaller than the Bohr radius $a_{B}$ is not very
important for the values of the electron energies. The effect is the
largest for $S$ levels, where the relative shift of energies goes like
$\left(R/a_{B}\right)^{2}\approx 10^{-10}$ where $R$ is nucleus radius
(in case of muonic atoms this shift is more important since it is
enhanced by $\left(m_{\mu}/m_{e}\right)^{2}\sim 10^{5}$).

Strong magnetic fields make the Coulomb problem essentially
one-dimensional. And in one space dimension a $1/|z|$ potential leads
to a spectrum unbounded from below: the ground state energy equals minus
infinity. The divergence of the potential at $z\to 0$ is regulated by
the Landau radius $a_{H}$: $\left|V(z)\right|\lesssim e^{2}/a_{H}$. It
follows from this consideration that the behaviour of the potential at
small distances determines the energy of the ground state. In this
section we will find how accounting for the finite size of the nucleus
modifies its electric potential.

An analytic formula for the Coulomb potential of a pointlike charge
along the $z$ axis screened by a magnetic field was derived in
\cite{2}: 
\begin{equation}
  \label{eq:0}
  \Phi(0,z)=\frac{e}{|z|}\left(1-e^{-|z|\sqrt{6m_{e}^{2}}}+e^{-\mu|z|}\right),
\end{equation}
where $z$ is the coordinate along the magnetic field,
$\mu\equiv\sqrt{6m_{e}^{2}+(2e^{3}B/\pi)}$, and the charge is 
located at the point $z=0$. The sum of the first two terms,
$\left(1-e^{-|z|\sqrt{6m_{e}^{2}}}\right)/|z|$, does not vary at
distances $z\ll 1/m_{e}$, and the smearing of the pointlike charge
along the $z$ axis within the domain with size $R\ll 1/m_{e}$ does not
affect it. At the  same time the last term,
$e^{-\mu|z|}/|z|$, becomes very sensitive
to the charge distribution for $B\gtrsim 1/(e^{3}R^{2}),~\mu R\gtrsim 1$.  

The potential in the plane transverse to the magnetic field ($z=0$)
was found in \cite{6} for $\mu\gg m_{e}$, $B\gg m_{e}^{2}/e^{3}$:
\begin{equation}
  \label{eq:0a}
  \Phi(\rho,0)=\frac{e}{\rho}\left(e^{-\mu\rho}+\frac{\sqrt{6m_{e}^{2}}}{\mu}\right),
\end{equation}
where $\vec \rho$ is the coordinate in the transverse plane. The
potential has a Yukawa behaviour both in the direction transverse to
the magnetic field and along the $z$ axis at distances $\rho,z\lesssim
l_{0}\equiv \frac{1}{\mu}\ln\frac{\mu}{\sqrt{6m_{e}^{2}}}$. At these
distances the potential is sensitive to the charge distribution. The
important question is whether the long range part of the potential
along the $z$ axis, $\left(1-e^{-|z|\sqrt{6m_{e}^{2}}}\right)/|z|$, is
also affected by the smearing of the pointlike charge in the
transverse plane. 

At $z\gg 1/m_{e}$ the potential $\Phi(\rho,z)$ has the following
behaviour \cite{6}:
\begin{eqnarray}
  \label{eq:0b}
  \Phi(\rho,z)=\frac{e}{\sqrt{z^{2}+\left(1+\frac{e^{3}B}{3\pi
          m_{e}^{2}}\right)}\rho^{2}}. 
\end{eqnarray}
The main contribution from the terms
$\left(1-e^{-|z|\sqrt{6m_{e}^{2}}}\right)/|z|$ to the value of the electron energy
comes from large distances, $1/m_{e}\ll z \ll 1/(e^{2}m_{e})$. To
significantly change this contribution the potential (\ref{eq:0b})
should noticeably differ from the Coulomb potential at distances $z\sim
1/(m_{e}e)$: 
\begin{eqnarray}
  \label{eq:0c}
  \frac{e^{3}B}{3\pi m_{e}^{2}}R^{2}\gtrsim
  \frac{1}{(em_{e})^{2}}\Rightarrow
  B\gtrsim \frac{3\pi}{e^{5}R^{2}}\approx 4\cdot 10^{10}~B_{0}.
\end{eqnarray}
We are not going to consider here such strong fields; so this effect
is neglected in what follows.

Thus, there are two parts in the potential: the first one,
$\left(1-e^{-|z|\sqrt{6m_{e}^{2}}}\right)/|z|$, does not depend on the charge
distribution inside the nucleus and the second one, originating from $e^{-\mu
  r}/r$ (where $r=\sqrt{\rho^{2}+z^{2}}$),
is determined by the charge distribution.

Another issue is the modification of the proton shape in a superstrong
magnetic fields which would lead to the variation of atomic
energies. As soon as the Landau radius of the  
electron becomes close to the proton radius the same happens with the
Landau radius of the proton. When the magnetic field further grows one
could expect that the size of the proton in the direction
perpendicular to the magnetic field shrinks. But the proton is not an
elementary particle and for $R\gg a_{H}$ the valence and sea quarks
will be oscillating inside domains with size of the order of
$a_{H}$ which are not necessarily situated at $\rho=0$. The
distribution of these rotating quarks inside the nucleus will be
defined by strong interactions, so there is no reason to think that
the nucleus in the transverse plane is squeezed down to
the size of $a_{H}$.\footnote{Nuclear core should prevent heavy ions
  from shrinking in the direction transverse to the magnetic field.}
What is really happening to the proton shape in
such a strong magnetic field is a subject for a separate study, while
here we will neglect the possible shrinking of the nucleus in
the magnetic field.

With the account of the specific features discussed above the potential
along the $z$ axis at $\rho\lesssim a_{H}$ has the following form: 
\begin{equation}
  \label{eq:1}
  \Phi(\rho,z)=
  \left\{
    \begin{array}{ll}
    \frac{e}{r}\left(1-e^{-r\sqrt{6m_{e}^{2}}}+h(R)e^{-\mu r}\right),&  
    r\geq R,\\
    \frac{e}{R}\left(1-e^{-R\sqrt{6m_{e}^{2}}}+h(r)e^{-\mu R}\right),&
    r< R,
  \end{array}
    \right.
\end{equation}
where $h(r)$ is determined by the charge distribution inside the nucleus.

Since the charge distribution inside the nucleus in such strong magnetic
field is not known, we will consider three following cases:
\begin{enumerate}
\item \underline{``Simple cut''} --- the potential outside the nucleus
  is equal to that of a pointlike charge and inside the nucleus it is
  constant and equals the value of the potential at the surface. This case
  corresponds to $h(r)=1$:
  \begin{equation}
  \label{eq:1cut}
  \Phi^{(1)}(\rho,z)=
  \left\{
    \begin{array}{ll}
    \frac{e}{r}
    \left(1-e^{-r\sqrt{6m_{e}^{2}}}+e^{-\mu r}\right),& 
    r\geq R,\\
    \frac{e}{R}\left(1-e^{-R\sqrt{6m_{e}^{2}}}+e^{-\mu R}\right),&
    r< R.
  \end{array}
    \right.
\end{equation}

\item \underline{Homogeneously charged sphere} --- taking into
  account that the last term $\varphi=eh(r)e^{-\mu R}/R$ in the expression
  (\ref{eq:1}) satisfies the Yukawa equation $\Delta
  \varphi-\mu^{2}\varphi=-4\pi Q(r)$ we obtain:
\begin{equation}
  \label{eq:1sphere}
  \Phi^{(2)}(\rho,z)=
  \left\{
    \begin{array}{ll}
    \frac{e}{r}\left(1-e^{-r\sqrt{6m_{e}^{2}}}+e^{-\mu r}
    \cdot\frac{1}{2\mu R}\left(e^{\mu R}-e^{-\mu R}\right)\right),&  
    r\geq R,\\
    \frac{e}{R}\left(1-e^{-R\sqrt{6m_{e}^{2}}}+e^{-\mu
      R}\cdot\frac{1}{2\mu r}\left(e^{\mu r}-e^{-\mu r}\right)\right),&
    r< R.
  \end{array}
    \right.
\end{equation}

\item \underline{Homogeneously charged ball} --- the potential can be
  easily found from the formula for a charged sphere: 
\begin{equation}
  \label{eq:1ball}
  \Phi^{(3)}(\rho,z)=
  \left\{
    \begin{array}{ll}
    \frac{e}{r}\left(1-e^{-r\sqrt{6m_{e}^{2}}}+e^{-\mu r}
    \cdot\frac{3}{2(\mu R)^{3}}\left(e^{\mu R}(\mu R-1)+e^{-\mu R}(\mu
      R+1)\right)\right),&  
    r\geq R,\\
    \frac{e}{R}\left(1-e^{-R\sqrt{6m_{e}^{2}}}+e^{-\mu R}
      \cdot\frac{3}{(\mu R)^{2}}\left(e^{\mu R}-
        \frac{(\mu R+1)}{2\mu r}\left(e^{\mu r}-e^{-\mu r}\right)\right)\right),&
    r< R.
  \end{array}
    \right.
\end{equation}
\end{enumerate}

\section{The finite size of the proton and  the hydrogen atomic levels}
\label{sec:finsize}

The following equation for the hydrogen atomic energies on which
the lowest Landau level (LLL) with $m=0$ splits was obtained in \cite{10}
by solving the Schr\"{o}dinger equation (and it was checked in \cite{6}
that the relativistic corrections can be neglected as far as the binding
energy is much smaller than the electron mass): 
\begin{equation}
2\ln \frac{z_0}{a_{B}}+\lambda+2\ln\lambda
+2\psi\left(1-\frac{1}{\lambda}\right)+4\gamma+2\ln
2=2\int\limits_{0}^{z_{0}}dz\int\frac{\left|R_{00}(\rho)\right|^{2}}{\sqrt{\rho^{2}+z^{2}}}d^{2}\rho\equiv I, 
\label{eq:2}
\end{equation}
where the energies of the atomic states are determined by $\lambda$,
$E\equiv -\left(m_{e}e^{4}/2\right)\lambda^{2}$, $a_{B}\equiv
1/(m_{e}e^{2})$ is the Bohr radius, $\psi$ is the logarithmic
derivative of the gamma function, and $\gamma=0.5772...$ is the
Euler's constant. $R_{00}(\rho)=e^{-\rho^{2}/4a_{H}^{2}}/\sqrt{2\pi
  a_{H}^{2}}$  is the wave function which corresponds to the transverse
motion of the electron occupying the ground ($n_{\rho}=m=0$) Landau
level. The dependence on the matching point $z_{0}$
cancels in (\ref{eq:2}) for $a_{H}\ll z_{0}\ll
1/\left(m_{e}e^{2}\right)$.

To take screening into account the factor $1/\sqrt{\rho^{2}+z^{2}}$ in
the right hand side of (\ref{eq:2}) should be substituted by:
\begin{equation}
  \label{eq:3}
  \frac{1}{\sqrt{\rho^{2}+z^{2}}}\rightarrow
  \frac{1}{\sqrt{\rho^{2}+z^{2}}}\left(1-e^{-\sqrt{\rho^{2}+z^{2}}\sqrt{6m_{e}^{2}}}+
    e^{-\sqrt{\rho^{2}+z^{2}}\sqrt{6m_{e}^{2}+2e^{3}B/\pi}}\right),
\end{equation}
which leads to the freezing of the values of atomic energies at $B\gg
m_{e}^{2}/e^{3}$ \cite{1,2}. To take the finite proton size into account
instead of (\ref{eq:3}) one should make the following substitution in
(\ref{eq:2}): 
\begin{equation}
  \label{eq:4}
  \frac{1}{\sqrt{\rho^{2}+z^{2}}}\rightarrow
  \Phi(\rho,z)/e,
\end{equation}
where $\Phi(\rho,z)$ is given by (\ref{eq:1}). For the right hand side
of (\ref{eq:2}) we get:
\begin{eqnarray}
\nonumber
  I&=&I_{1}+I_{2}+I_{3}\equiv 2\int\limits_{0}^{R}dz\int\limits_{0}^{\sqrt{R^{2}-z^{2}}}2\pi\rho
  d\rho\left|R_{00}(\rho)\right|^{2}
  \frac{1}{R}\left[1-e^{-R\sqrt{6m_{e}^{2}}}+h(\sqrt{\rho^{2}+z^{2}})e^{-\mu
      R}\right]+\\ 
  &+&2\int\limits_{0}^{R}dz\int\limits_{\sqrt{R^{2}-z^{2}}}^{\infty}2\pi\rho
  d\rho\left|R_{00}(\rho)\right|^{2}\frac{1}{\sqrt{\rho^{2}+z^{2}}}
  \left[1-e^{-\sqrt{\rho^{2}+z^{2}}\sqrt{6m_{e}^{2}}}+h(R)e^{-\mu\sqrt{\rho^{2}+z^{2}}}\right]+
  \nonumber\\
  &+&2\int\limits_{R}^{z_{0}}dz\int\limits_{0}^{\infty}2\pi\rho
  d\rho\left|R_{00}(\rho)\right|^{2}\frac{1}{\sqrt{\rho^{2}+z^{2}}}
  \left[1-e^{-\sqrt{\rho^{2}+z^{2}}\sqrt{6m_{e}^{2}}}+
    h(R)e^{-\mu\sqrt{\rho^{2}+z^{2}}}\right].\;
  \label{eq:5}
\end{eqnarray}

The sum $I_{1}+I_{2}$ for $a_{H}\approx R$ is of order $1$ and
can be safely neglected. For $a_{H}\ll R$ we have:
\begin{eqnarray}
  \label{eq:5a}
  I_{1}+I_{2}\approx
  2R\cdot\frac{1}{R}\left(1-e^{-R\sqrt{6m_{e}^{2}}}\right)+
  \frac{2e^{-\mu R}}{R}\int_{0}^{R}h(z)dz\equiv
  2\left(1-e^{-R\sqrt{6m_{e}^{2}}}+ \langle h\rangle e^{-\mu R}\right).\;
\end{eqnarray}
We see that for three examples considered in Sec. \ref{sec:potential} 
the sum $I_{1}+I_{2}$ rapidly diminishes when $B$ grows, so we can
safely neglect these terms.\footnote{Let us note that the term
  $\langle h\rangle e^{-\mu R}$ is not necessarily decreasing with
  the magnetic field because $h(r)$ could correspond to any charge
  distribution inside the nucleus including a pointlike
  distribution. In case of a pointlike distribution this term  would
  prevent the ground energy level from going up. 
}

For $I_{3}$ we have:
\begin{eqnarray}
I_{3}&\approx&
2\int\limits_{\sqrt{R^{2}+a_{H}^{2}}}^{z_{0}}\frac{dz}{z}\left[1-e^{-z\sqrt{6m_{e}^{2}}}+
  h(R)e^{-\mu z}\right]
\approx\nonumber\\
&\approx&
2\left[\ln\frac{z_{0}}{\sqrt{R^{2}+a_{H}^{2}}}-E_{1}\left(\sqrt{R^{2}+a_{H}^{2}}\sqrt{6m_{e}^{2}}\right)+
h(R)E_{1}\left(\mu\sqrt{R^{2}+a_{H}^{2}}\right)\right],
  \label{eq:6}
\end{eqnarray}
where
\begin{eqnarray}
  E_{1}(x)&\equiv& \int\limits_{x}^{\infty}\frac{e^{-t}}{t}dt, \nonumber\\
  \left. E_{1}(x)\right|_{x\ll 1}&=&-\gamma-\ln
  x-\sum\limits_{n=1}^{\infty}\frac{(-1)^{n}x^{n}}{n\cdot n!},\nonumber\\
  \left. E_{1}(x)\right|_{x\gg  1}&=&\frac{e^{-x}}{x}\left(1-\frac{1}{x}+\frac{1\cdot
    2}{x^{2}}-\frac{1\cdot 2\cdot 3}{x^{3}}+\dots\right).
  \label{eq:7}
\end{eqnarray}

Substituting (\ref{eq:6}) in (\ref{eq:2}) we get  an equation which
determines the values of $\lambda$ and the corresponding values of
the energies on which the Landau level with $n_{\rho}=m=0$ splits by the
screened Coulomb potential: 
\begin{eqnarray}
&&\ln \frac{z_0}{a_{B}}+\frac{\lambda}{2}+\ln\lambda
+\psi\left(1-\frac{1}{\lambda}\right)+2\gamma+\ln 2=\nonumber\\
&&\ln\frac{z_{0}}{\sqrt{R^{2}+a_{H}^{2}}}-E_{1}\left(\sqrt{R^{2}+a_{H}^{2}}\sqrt{6m_{e}^{2}}\right)+
h(R)E_{1}\left(\mu\sqrt{R^{2}+a_{H}^{2}}\right).
\label{eq:8z}
\end{eqnarray}

The dependence on the matching point $z_{0}$ cancels and finally we
obtain the equation which determines the values of the freezing atomic
energies with the account of the finite proton size: 
\begin{eqnarray}
&&\frac{\lambda}{2}+\ln\lambda
+\psi\left(1-\frac{1}{\lambda}\right)+2\gamma+\ln 2=\nonumber\\
&&\ln\frac{a_{B}}{\sqrt{R^{2}+a_{H}^{2}}}-E_{1}\left(\sqrt{R^{2}+a_{H}^{2}}\sqrt{6m_{e}^{2}}\right)+
h(R)E_{1}\left(\mu\sqrt{R^{2}+a_{H}^{2}}\right).
\label{eq:8}
\end{eqnarray}

In the limit $B\gg 1/\left(e^{3}R^{2}\right)$ the right hand side of
(\ref{eq:8}) does not depend on $R$ and we obtain: 
\begin{eqnarray}
  \label{eq:9}
  &&\left. I_{3}\right|_{B\gg 1/\left(e^{3}R^{2}\right)}=2\left(\ln
    z_{0}\sqrt{6m_{e}^{2}}+\gamma\right),\\ 
  &&\lambda^{\rm lim}+2\ln \lambda^{\rm
    lim}+2\psi\left(1-\frac{1}{\lambda^{\rm lim}}\right)
  +2\gamma+2\ln 2=\ln\left(\frac{6}{e^{4}}\right).
  \label{eq:9a}
\end{eqnarray}
There are two ways to satisfy the equation (\ref{eq:9a}) which has the
large logarithm in the right hand side. The first one is to take a large
$\lambda^{\rm lim}$ which will correspond to the ground energy level. The
second one is to choose $\lambda^{\rm lim}$ close to the poles of
$\psi(1-\frac{1}{\lambda^{\rm lim}})$. The logarithmic derivative of
the gamma function $\psi(x)$ has poles at $x=0,-1,-2\dots$ which
defines a series of $\lambda^{\rm lim}\approx 1/n,~n=1,2\dots$ This
tower of $\lambda^{\rm lim}$ corresponds to the well known Balmer
series of hydrogen atomic energies $E_{n}^{\rm lim}\approx -(m_{e}e^{4})/(2n^{2})$.

For the ground level from (\ref{eq:9a}) we obtain $\lambda^{\rm lim}=6.9$ instead of the
value for a pointlike proton $\lambda^{\rm lim}=11.2$ obtained in
\cite{2}. Let us note that for $B\gg 1/(e^{3}R^{2})$ all our
approximations have a very good accuracy, so this result should have
a good accuracy as well. To check it we solved the Schr\"{o}dinger
equation numerically. Since the adiabatic approximation is applicable
($a_{H}\ll a_{B},~B\gg m_{e}^{2}e^{3}$) one has to solve the one
dimensional Schr\"{o}dinger equation with an effective potential $\bar V(z)$:
\begin{equation}
\frac{d^2\chi}{dz^2} + 2m_{e}(E-\bar V)\chi = 0 \;\; ,
\label{eq:010}
\end{equation}
\begin{eqnarray*}
E \equiv -\frac{m_{e}e^{4}}{2}\lambda^{2} \; , \;\;
\bar V(z)\equiv-\frac{e}{a_H^2}\int\limits_0^\infty \Phi(\rho,z)
\exp\left(-\frac{\rho^2}{2a_H^2}\right)\rho d\rho.
\end{eqnarray*}

In order to numerically solve equation (\ref{eq:010}) an analytical
formula for the averaged potential energy $\bar V(z)$ is needed. 
For $B\gg 1/(eR^{2})$, $a_{H}\ll R$, the potentials
(\ref{eq:1cut})--(\ref{eq:1ball}) do not vary with $\rho$ for
$\rho\lesssim a_{H}$, so the averaging over the transverse direction does
not change them, i.e. $\bar V^{(i)}(z)\approx
-e\Phi^{(i)}(0,z),~i=1,2,3$. For $B\ll 1/(eR^{2})$, $a_{H}\gg R$, the
modification of the potential at distances $r<R$ does not affect the
electron motion while the averaging is very important since it removes
$1/r$ singularity of $\Phi(\rho,z)$. So we need an
analytical formula for the averaged potential energy for intermediate
fields, $B\sim 1/(eR^{2})$ at which $a_{H}\sim R$. However at
distances $r\sim R\sim a_{H}$ the screening does not occur (because
$1/\mu\approx 10a_{H}$) so for $B\sim 1/(eR^{2})$ one should average
the non-screened potential at these distances.

Without taking screening into account the potential energy of the
electron in the external electric potential of a homogeneously charged
sphere has the following form:
\begin{eqnarray}
  \label{eq:011}
  \Phi^{(0)}(\rho,z)=\left\{
    \begin{array}{ll}
    \frac{e}{\sqrt{\rho^{2}+z^{2}}},&\sqrt{\rho^{2}+z^{2}}\geq R,\\
    \frac{e}{R},&\sqrt{\rho^{2}+z^{2}}< R.
  \end{array}
  \right.
\end{eqnarray}

The formula for the corresponding averaged potential energy $\bar
V^{(0)}(z)$ looks like:
\begin{eqnarray}
  \label{eq:012}
  \bar V^{(0)}(z)=\left\{
  \begin{array}{ll}
    -e^{2}\left(\frac{1}{R}\left(1-e^{(z^{2}-R^{2})/2a_{H}^{2}}\right)+
    \frac{1}{a_{H}}\sqrt{\frac{\pi}{2}}e^{z^{2}/2a_{H}^{2}}{\rm erfc}\left(\frac{R}{a_{H}\sqrt{2}}\right)\right),&|z|<R,\\
    -e^{2}\frac{1}{a_{H}}\sqrt{\frac{\pi}{2}}e^{z^{2}/2a_{H}^{2}}{\rm
      erfc}\left(\frac{|z|}{a_{H}\sqrt{2}}\right),&|z|\geq R,
  \end{array}\right.
\end{eqnarray}
where ${\rm erfc}(x)$ is the complementary error function:
\begin{eqnarray}
  \label{eq:012a}
  {\rm erfc}(x)\equiv 1-{\rm erf}(x)=\frac{2}{\sqrt{\pi}}\int\limits_{x}^{\infty}e^{-y^{2}}dy.
\end{eqnarray}

The extension of the formula (\ref{eq:012}) to the entire range of
distances and magnetic fields for the potential energies $\bar V^{(1)}(z)$ and
$\bar V^{(2)}(z)$ which correspond to the potentials $\Phi^{(1)}(\rho,z)$ and
$\Phi^{(2)}(\rho,z)$ is given by ($i=1,2$):
\begin{eqnarray}
  \label{eq:012cut}
  \bar V^{(i)}(z)=\left\{
  \begin{array}{ll}
    -e^{2}S_{1}^{(i)}\left(\frac{1}{R}\left(1-e^{(z^{2}-R^{2})/2a_{H}^{2}}\right)+
    \frac{1}{a_{H}}\sqrt{\frac{\pi}{2}}e^{z^{2}/2a_{H}^{2}}{\rm erfc}\left(\frac{R}{a_{H}\sqrt{2}}\right)\right),&|z|<R,\\
    -e^{2}S_{2}^{(i)}\frac{1}{a_{H}}\sqrt{\frac{\pi}{2}}e^{z^{2}/2a_{H}^{2}}{\rm
      erfc}\left(\frac{|z|}{a_{H}\sqrt{2}}\right),&|z|\geq R;
  \end{array}\right.\\
S_{1}^{(1)}=\left(1-e^{-R\sqrt{6m_{e}^{2}}}+e^{-\mu R}\right),~ S_{2}^{(1)}=\left(1-e^{-|z|\sqrt{6m_{e}^{2}}}+e^{-\mu
        |z|}\right);\\
S_{1}^{(2)}=\left(1-e^{-R\sqrt{6m_{e}^{2}}}+e^{-\mu
    R}\cdot\frac{1}{2\mu |z|}\left(e^{\mu |z|}-e^{-\mu
      |z|}\right)\right),~\\
S_{2}^{(2)}=\left(1-e^{-|z|\sqrt{6m_{e}^{2}}}+e^{-\mu
         |z|}\cdot\frac{1}{2\mu R}\left(e^{\mu R}-e^{-\mu
           R}\right)\right).
\end{eqnarray}

The formula (\ref{eq:012cut}) has the correct
behaviour both for $B\ll 1/(e^{3}R^{2})$, $a_{H}/e\gg R$, (because
the screening factors $S^{(i)}_{1}\approx 1$ and $S^{(i)}_{2}\approx 1$ for $|z|<R$) and for $B\gg
1/(eR^{2})$, $a_{H}\ll R$, (because averaging does not change $\Phi^{(0)}(\rho,z)$: 
$\bar V^{(0)}(z)|_{B\gg 1/(eR^{2})}\approx -e\Phi^{(0)}(0,z)$). Since
these ranges of magnetic fields overlap the formula (\ref{eq:012cut}) is correct for
all $B$.

\begin{figure}[b]
  \centering
  \includegraphics[scale=0.65]{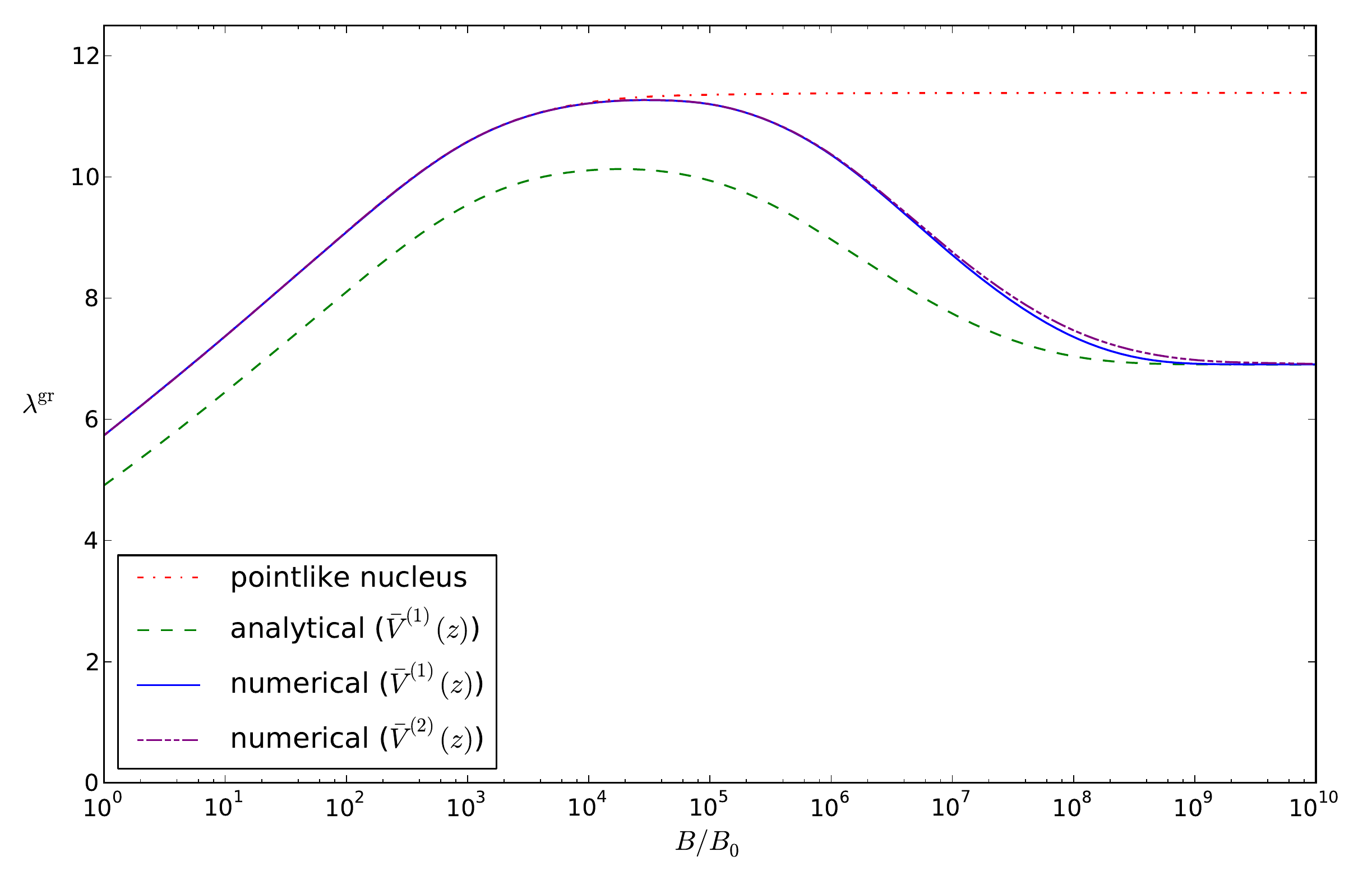}
  \caption{The dependence of $\lambda^{\rm gr}$ on the magnetic field;
    the ground state energy equals $E_{0}\equiv
    -(m_{e}e^{4}/2)\left(\lambda^{\rm gr}\right)^{2}$. The dot-dashed 
    (red) line corresponds to the pointlike nucleus, the dashed
    (green) line --- to the analytical formula (\ref{eq:8}), the 
    solid (blue) and the dashed line with two dots (purple) --- to the
    numerical solutions of (\ref{eq:010}) with $\bar V^{(1)}(z)$ and
    $\bar V^{(2)}(z)$ correspondingly.}
  \label{fig:1}
\end{figure}

In Fig. \ref{fig:1} the behaviour of $\lambda^{\rm gr}$ (which
corresponds to the ground level) according to the
analytical expression (\ref{eq:8}) for $h(r)=1$ (``Simple cut'') and
according to the results of the numerical solution of
Eq. (\ref{eq:010}) with $\bar V^{(1)}(z)$ and $\bar V^{(2)}(z)$  are
shown. We see how $\lambda^{\rm gr}$ is  
going down (or the ground level is going up) when the Landau radius
$a_{H}$ becomes of the order of the proton radius ($B\sim
1/(eR^{2})\approx 2\cdot 10^{5}~B_{0}$). The raising stops 
and the energy freezes at $B\sim 1/(e^{3}R^{2})\approx 3\cdot
10^{7}B_{0}$.  According to Fig. \ref{fig:1} $\bar V^{(1)}(z)$
and $\bar V^{(2)}(z)$ lead to practically the same dependence of
$\lambda^{\rm gr}$ on $B$ so we use the potential $\bar V^{(1)}(z)$ in what follows. 
\footnote{The potential
  $\Phi^{(3)}(\rho,z)$ varies inside the nucleus even in 
the non-screened case, that is why we do not have an analytical formula for
the averaged potential energy. Nevertheless we made an estimate
for $\bar V^{(3)}(z)$ and found the numerical results for
$\lambda^{\rm gr}$ which appeared to be rather close to the results
obtained with $\bar V^{(1)}(z)$ and $\bar V^{(2)}(z)$.}

To present a qualitative explanation of the phenomenon of the rising
of the ground energy level let us put $z_{0}\approx a_{B}$ in
Eq. (\ref{eq:2}). In case of a pointlike charge the 
main contributions to $I$ (the right hand side of (\ref{eq:2})) for $B\gg
m_{e}^{2}/e^{3}$ come from integrating over $a_{H}<|z|<1/\mu$ and
$1/m_{e}<|z|<z_{0}\approx a_{B}$ where the potential has the Coulomb ($1/|z|$) behaviour:
\begin{eqnarray}
  \label{eq:Ipointlike}
  \left.I\right|_{eB<1/R^{2}}\approx
  \ln\frac{1}{a_{H}^{2}\mu^{2}}+\ln\left(m_{e}^{2}a_{B}^{2}\right)
  \approx \ln\frac{1}{e^{2}}+\ln\frac{1}{e^{4}}=\ln\frac{1}{e^{6}}.
\end{eqnarray}
When the magnetic field grows the Landau radius $a_{H}$ approaches the
proton radius $R$ and the first logarithm in (\ref{eq:Ipointlike})
should be substituted by $\ln\left(1/(R\mu)^{2}\right)$ since at
$|z|<R$ the potential does not have a $1/|z|$ behaviour. When $B$
further grows and $e^{3}B$ approaches $1/R^{2}$ the first logarithm in 
(\ref{eq:Ipointlike}) goes away:
\begin{eqnarray}
  \label{eq:Ifinsize}
  \left.I\right|_{e^{3}B\approx \mu^{2}>1/R^{2}}\approx
  \ln\left(m_{e}^{2}a_{B}^{2}\right)=\ln\frac{1}{e^{4}}.
\end{eqnarray}
The decreasing of $I$ corresponds to the diminishing of $\lambda$ and,
therefore, the ground energy level goes up.

\section{Critical nucleus charge}
\label{sec:finsizeZ}

It is well know that the Dirac equation in a pointlike Coulomb
potential is not self-consistent for an electric charge $Z>137$. If
the finite size of the nucleus is taken into account then the Dirac
equation becomes self-consistent. For $Z\approx 172$ the ground energy
level reaches the lower continuum, $\varepsilon=-m_{e}$, and two
electron--positron pairs are created from the vacuum \cite{3}. The
electrons occupy the ground energy level while the positrons become
free (two pairs are created due to the spin degeneracy of the ground
energy level). This is known as the critical nucleus charge
phenomenon.  

In the presence of a magnetic field the value of the critical nucleus
charge diminishes \cite{4}. The atomic energies were found by
solving the Dirac equation. In \cite{4} it was transformed into a
set of two one-dimensional differential equation of first order which
can be done when the adiabatic approximation is valid ($B\gg 
B_{0}\left(Ze^{2}\right)^{2}$):  
\begin{equation}
  \label{eq:100}
  \begin{array}{c}
  g_{z}-(\varepsilon+m_{e}-\bar{V})f=0,\\  
  f_{z}+(\varepsilon-m_{e}-\bar{V})g=0,
\end{array}
\end{equation}
where $g_{z}\equiv dg/dz,~f_{z}\equiv dh/dz$; $\varepsilon$ is the
energy eigenvalue of the Dirac equation; the bispinor
$\psi_{e}=\left(\varphi_{e}\atop\chi_{e}\right)$ of the electron is
decomposed into $\varphi_{e}=\left(0\atop
  g(z)\exp\left(-\rho^{2}/4a_{H}^{2}\right)\right)$,
$\chi_{e}=\left(0\atop
  if(z)\exp\left(-\rho^{2}/4a_{H}^{2}\right)\right)$. The averaged
potential energy $\bar V(z)$ is defined in the same way as it was done in 
Sec. \ref{sec:finsize}.

The analytical formula which describes the dependence of the atomic
energies $\varepsilon$ on $B$ was derived in \cite{4}: 
\begin{eqnarray}
 Ze^2\ln\left(2\frac{\sqrt{m_e^2-\varepsilon^2}}{\sqrt{eB}}\right) +
\arctan\left(\sqrt{\frac{m_e+\varepsilon}{m_e-\varepsilon}}\right) 
+\arg\Gamma\left(-\frac{Ze^2 \varepsilon}{\sqrt{m_e^2 -
\varepsilon^2}} + iZe^2\right)&& \nonumber\\*-  \arg \Gamma(1+2iZe^2) -
\frac{Ze^2}{2}(\ln 2 + \gamma) =\frac{\pi}{2} &+& n\pi \;\; , \label{eq:100a}
\end{eqnarray}
where the argument of the gamma function is given by 
\begin{equation}
\arg\Gamma(x+iy) = -\gamma y + \sum_{k=1}^\infty\left(\frac{y}{k}
- \arctan\frac{y}{x+k-1}\right) \;\; . \label{eq:100arg}
\end{equation}
For the ground level at $\varepsilon > 0$ one should take $n=0$, while
for $\varepsilon < 0$ it should be changed to $n=-1$.

Substituting $\varepsilon=-m_{e}$ into (\ref{eq:100a}) the formula for
the critical nucleus charge in a magnetic field was found in \cite{4}:
\begin{equation}
  \label{eq:100bmax}
  \frac{B}{B_{0}}=2(Z_{cr}e^{2})^{2}\exp\left(-\gamma+\frac{\pi-2\arg\Gamma
        (1+2iZ_{cr}e^{2})}{Z_{cr}e^{2}}\right).
\end{equation}
According to this formula the critical nucleus charge diminishes with
the magnetic field and for $B\approx 10^{2}B_{0}$ the uranium becomes
critical (equations (\ref{eq:100a}) and (\ref{eq:100bmax}) are valid for
$B>\max\left\{\left(Ze^{2}\right)^{2}B_{0},B_{0}/\left(Ze^{2}\right)^{2}\right\}$).

To satisfy the matching condition used to derive the formula
(\ref{eq:100a}) the potential should be Coulomb at distances
$l>z_{0}$, where $z_{0}\ll Ze^{2}/(2m_{e})$ is the matching point (see 
\cite{4} for details). This condition is violated for the screened
potential. Since we do not have an analytical formula for the energy
levels which takes screening into account in the relativistic case, in
\cite{6} we solved the Dirac equation numerically.  In order to do
this we followed the paper \cite{4} where the system (\ref{eq:100}) was
transformed into one second order differential equation for $g(z)$. By
substituting $g(z)=\left(\varepsilon+m_{e}-\bar{V}\right)^{1/2}\chi(z)$ 
a Schr\"{o}dinger-like equation for the function $\chi(z)$ was obtained
in \cite{4}:\footnote{This method of reduction of the Dirac equation
  to a Schr\"{o}dinger-like equation was originally proposed by
  V.S. Popov to analyze the critical nucleus charge phenomenon
  qualitatively.} 
\begin{equation}
\frac{d^2\chi}{dz^2} + 2m_{e}(E-U)\chi = 0 \;\; , \label{eq:102}
\end{equation}
\begin{eqnarray*}
E = \frac{\varepsilon^2 - m_e^2}{2m_e} \; , \;\;
U = \frac{\varepsilon}{m_e}\bar V- \frac{1}{2m_e}\bar V^2
+ \frac{\bar V''}{4m_e(\varepsilon + m _e-\bar V)} + \frac{3/8(\bar
V')^2}{m_e(\varepsilon +m_e - \bar V)^2} \; .
\end{eqnarray*}
In \cite{6} we used the equation (\ref{eq:102}) for numerical calculations. The
numerical results for the pointlike nucleus were presented in \cite{6}
where the freezing of the ground energy level was obtained in the
relativistic domain and the values of the critical magnetic fields were
found for $Z\geq 50$. Nuclei with $Z<50$ do not become critical due to
screening.

\begin{figure}[t]
  \centering
  \includegraphics[scale=0.5]{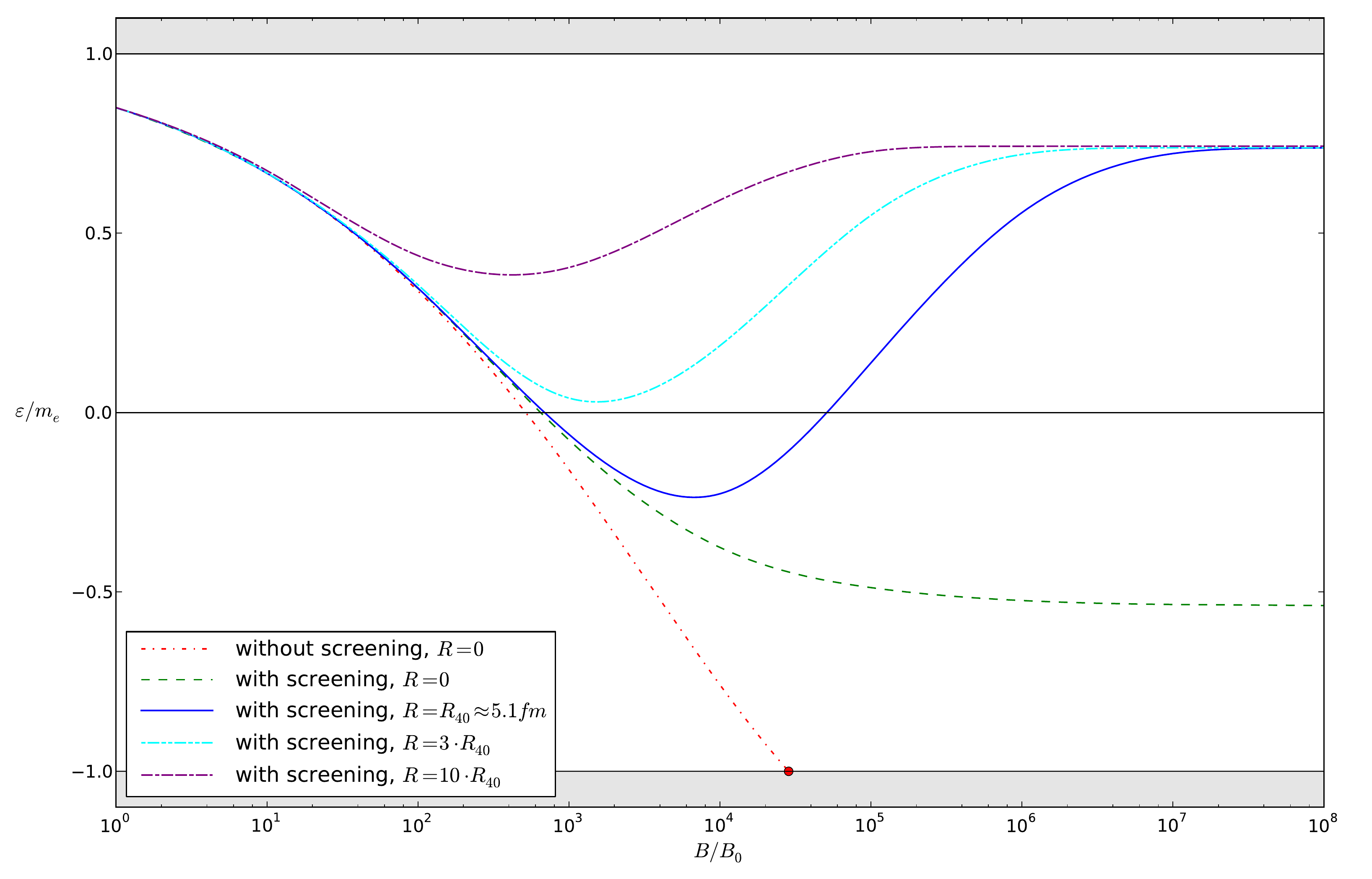}
  \caption{The dependence of the ground energy level on the magnetic field
    for $Z=40$. The dot-dashed (red) line corresponds to a pointlike
    potential without screening, the dashed (green) line --- to a
    pointlike potential with screening, the solid (blue) line --- to
    a potential which takes into account both screening and the finite
    size of the nucleus $R=R_{40}\approx 5.1$ fm. Cyan and purple lines
    (linestyles are shown in legend) corresponds to (hypothetical)
    larger nucleus radii, $R=3R_{40}$ and $R=10 R_{40}$.} 
  \label{fig:Z40}
\end{figure}

Our next step is to generalize the results of \cite{6} in order to take
the finite nucleus size into account. Let us note that in most cases
the averaged potential energy $\bar V$ is smooth even when the 
potential $\Phi (\rho,z)$ has a cusp at the nucleus boundary. However
due to our approximations in averaging  $\bar V^{(1)}(z)$ done in the
Sec. \ref{sec:finsize} the potential does have 
this cusp which is quite significant for $B\gg 1/(eR^{2})$. It means
that we have a $\delta$-singularity in the effective potential $U$ and
this should be taken into account in numerical calculations.

Substituting $\bar V^{(1)}(z)$ for $\bar V(z)$ we checked that the
equation (\ref{eq:102}) gives the values of the binding energy for
hydrogen $E=\frac{\varepsilon^{2}-m_{e}^{2}}{2m_{e}}\equiv
-\frac{m_{e}e^{4}}{2}\lambda^{2}$, which is very close to the ones
obtained from the Schr\"{o}dinger equation. 

Substituting the nucleus radius $R_{Z}=r_{0}A^{1/3}$ and $\bar V(z)=Z\bar
V^{(1)}(z)$ into (\ref{eq:102}) we numerically calculate the
value of the ground  state energy of the hydrogen-like ion with charge
$Z$.\footnote{In the numerical calculations we are using $r_{0}=1.1 fm$ and $A=2.5Z$.} In
Fig. \ref{fig:Z40} the dependence of the ground state energy 
on $B$ is shown for $Z=40$. We observe the rising of the ground energy
level in the relativistic domain. The curves for 
a nucleus with $Z=40$ and radii $3R_{40}$ and $10 R_{40}$ are plotted to check
how the energy depends on nucleus radius. We see that the
rising starts at $B\sim 1/(eR^{2})$ and stops at $B\sim
1/(e^{3}R^{2})$.  The limiting energy does not depend on $R$.

\begin{figure}[t]
  \centering
  \includegraphics[scale=0.5]{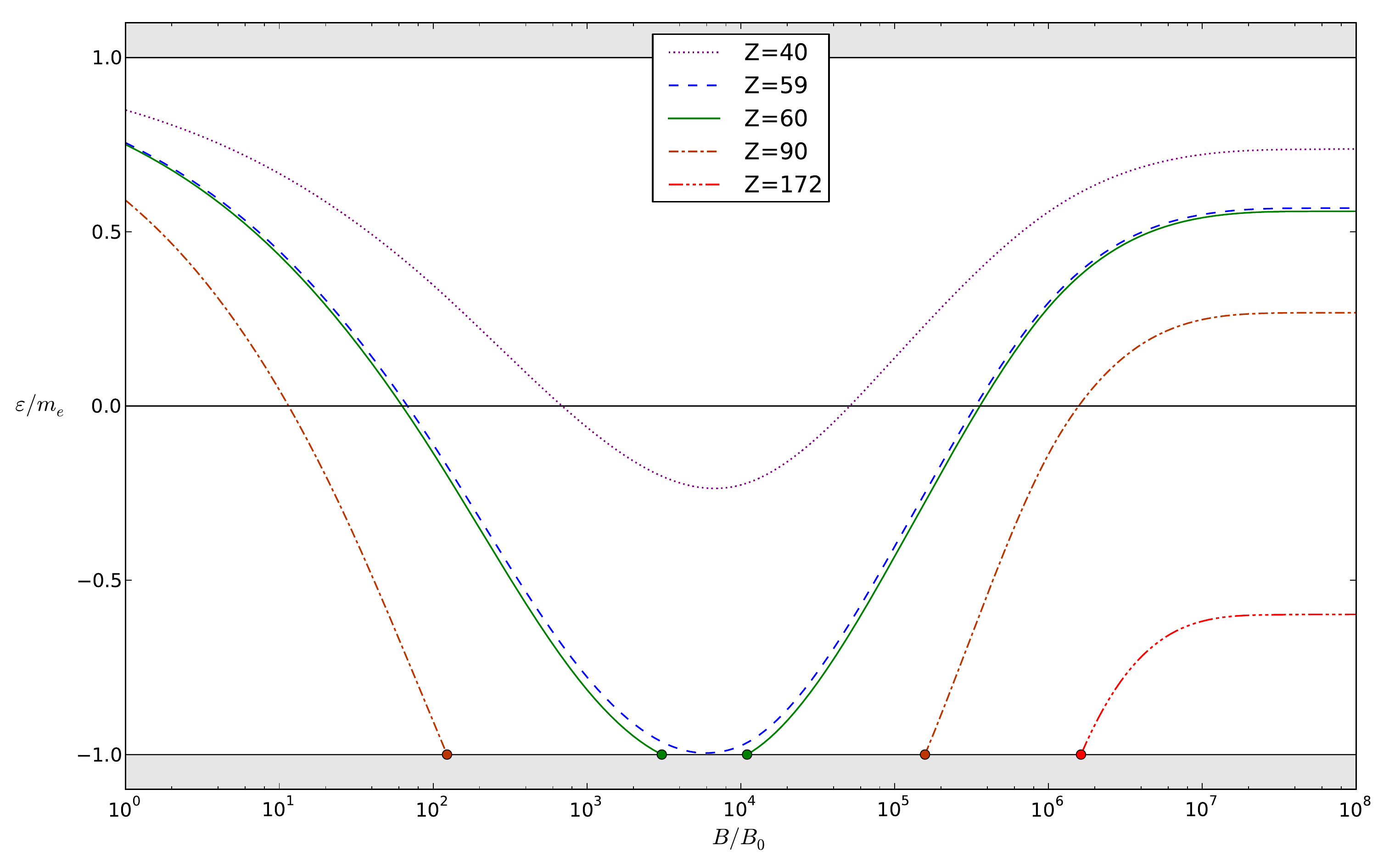}
  \caption{The dependence of the ground energy level on magnetic field
    for $Z=40,59,60,90,172$. The correspondence between
    charge $Z$ and color (linestyle) is shown in the legend.}
  \label{fig:levels}
\end{figure}

In Fig. \ref{fig:levels} the dependence of the ground state energy on
the magnetic field for $Z=40,59,60,90,172$ is shown. We see that ions with
$Z<60$ never become critical while a nucleus with $Z=60$ is critical
only within the small range of magnetic fields around $B\approx
10^{4}B_{0}$. For larger $Z$ the range of magnetic fields in which
ions are critical becomes wider. Ions become critical at $B$ a little bit
larger than the critical field for a pointlike nucleus. The
rising of the ground energy level makes the ions noncritical for
strong enough magnetic fields. Even an ion with $Z=172$ becomes
noncritical for $B>1.6\cdot 10^{6}B_{0}$ while it is critical for $B=0$. 
To estimate the value of the nucleus charge at which the ``final'', or
``second'', freezing energy reaches the lower continuum we demanded that
the nucleus should be critical for $B=10^{8}B_{0}$ and found that it
is satisfied for $Z\geq 210$. It means that only ions with $Z\geq
210$ remains critical regardless of the value of the magnetic fields.

\begin{figure}[t]
  \centering
  \includegraphics[scale=0.5]{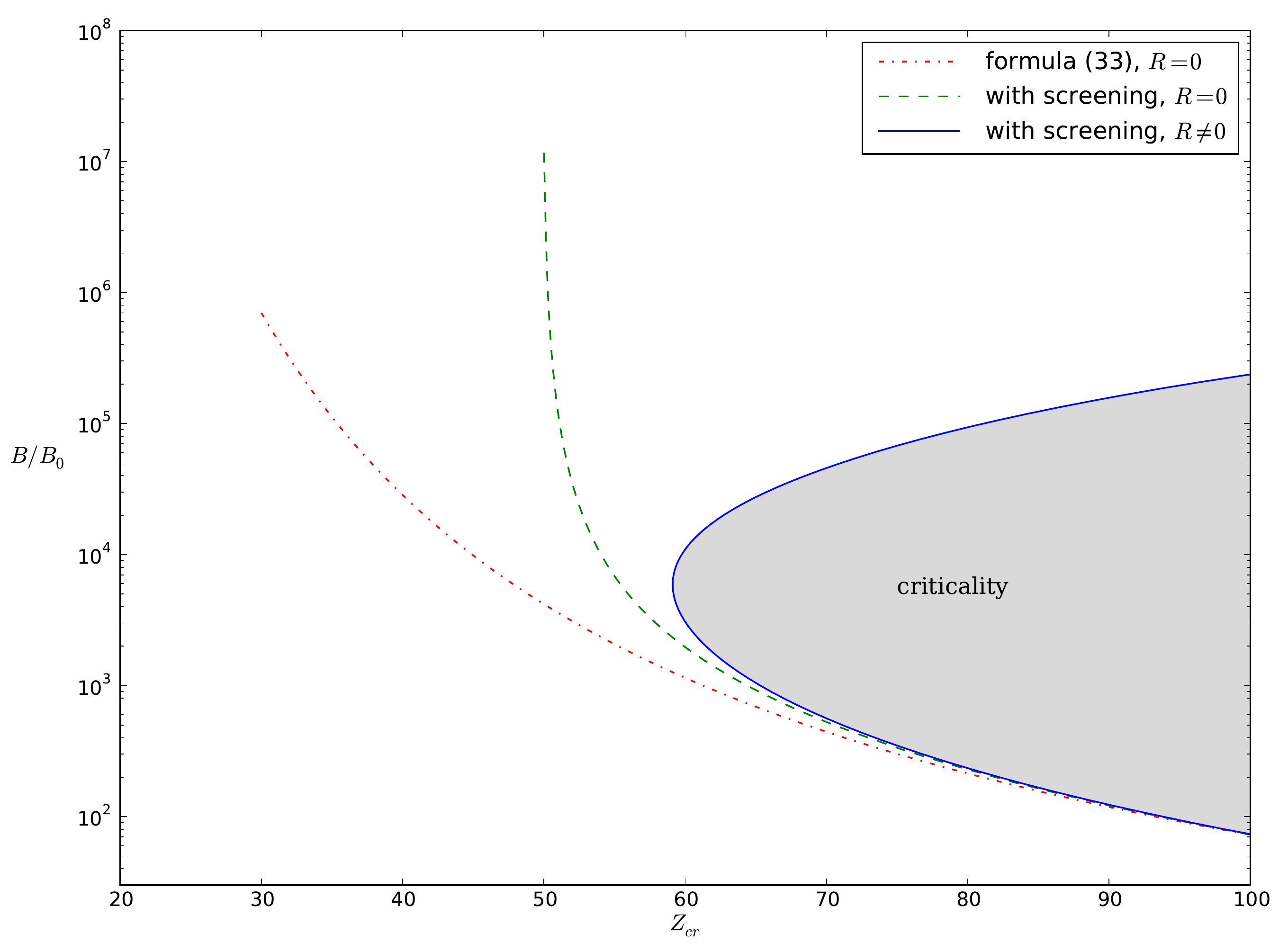}
  \caption{The values of magnetic fields at which the nuclei with
    charge $Z_{cr}$ becomes critical: a) without screening according
    to eq. (\ref{eq:100bmax}), dot-dashed (red) line; b) numerical
    results with screening for pointlike nucleus, dashed (green) line;
    c) numerical results which take finite size of the nucleus into
    account, solid (blue) line.}
  \label{fig:Bmax}
\end{figure}

In Fig. \ref{fig:Bmax} the dependence of critical nucleus charge on
$B$ is shown.

\section{The hydrogen atomic levels in a superstrong $B$ and the
  proton motion}
\label{sec:finmass}

The two-body problem in the presence of a homogeneous magnetic field
constant in time was analyzed in the papers \cite{7,7a,8} where it was
found that for an electrically neutral system the separation of the
center-of-mass and relative coordinates can be carried out
explicitly. According to Eq. (12) from \cite{8} the Hamiltonian which
describes the relative motion of the electron and the proton in an
external magnetic field $\vec B$ looks like:
\begin{eqnarray}
  \label{eq:401}
  \hat H_{rel}=\frac{\vec K^{2}}{2M}+\frac{e}{M}\left(\vec K\times\vec
  B\right)\vec r+\frac{1}{2m_{r}}\vec
p^{2}+\frac{e}{2}\left(\frac{1}{m_{e}}-\frac{1}{m_{p}}\right)\vec
B\left(\vec r\times\vec p\right)+\\ \nonumber \frac{e^{2}}{8m_{r}}\left(\vec
  B\times\vec r\right)^{2}+V(r),
\end{eqnarray}
where the momentum $\vec K$ is an eigenvalue of the generalized
momentum operator responsible for the motion of the center-of-mass of
the atom, $\vec r\equiv \vec r_{e}-\vec r_{p}$ and $\vec p\equiv
-i \partial/\partial\vec r$ are the relative coordinate and momentum of
the electron and proton, $M=m_{e}+m_{p},~m_{r}\equiv m_{e}m_{p}/M$ is the
reduced mass, and $V(r)$ describes the Coulomb interaction between
the electron and the proton. The screening of the Coulomb interaction
in a superstrong $\vec B$ should be taken into account as well \cite{1,2}.

Let us consider an atom at rest ($\vec K=0$) in a strong magnetic field
$B\gg m_{e}^{2}e^{3}$ in which the adiabatic approximation is
applicable. The relative motion of the electron and proton in the
direction of the magnetic field is determined by the potential $V(z)$,
while in the direction transverse to the magnetic field it is
determined by $\vec B$.

Neglecting in a first approximation the Coulomb attraction for
the atomic energy levels from (\ref{eq:401}) we obtain:
\begin{eqnarray}
  \label{eq:402}
  E_{n_{\rho}m}=\frac{eB}{m_{e}}\left(n_{\rho}+\frac{|m|+m+1}{2}\right)+
  \frac{eB}{m_{p}}\left(n_{\rho}+\frac{|m|-m+1}{2}\right).
\end{eqnarray}

In what follows we will be interested in the states of the hydrogen atom
which originate from the lowest Landau level (LLL), for which
$n_{\rho}=0,~m=0,-1,-2,\dots$ and the electron spin is antiparallel to
$\vec B$. For LLL the contribution of the first term in (\ref{eq:402})
summed with the energy of interaction of the electron spin with
the external magnetic field $\vec B$ is zero while the second term
gives a nonzero contribution. Thus the energies of the atomic states with
different $m$ are shifted by
\begin{eqnarray}
  \label{eq:403}
  \Delta E_{m}=\frac{eB}{m_{p}}|m|.
\end{eqnarray}

Taking into account the motion along the $z$-axis governed by the (screened)
Coulomb potential we obtain the following expression for the atomic
energies:
\begin{eqnarray}
  \label{eq:404}
  E=-\frac{m_{r} e^{4}}{2}\lambda^{2}+\frac{eB|m|}{m_{p}}\equiv
  E_{\lambda}+\frac{eB|m|}{m_{p}},
\end{eqnarray}
where in the first term $m_{e}$ can be safely substituted for the reduced
mass $m_{r}$ since the numerical difference between $m_{r}$ and $m_{e}$ is
very small. The values of $\lambda$ are determined by the following
transcendental equation (see \cite{2},Eq. (57)):
\begin{eqnarray}
  \label{eq:405}
  \ln\left(\frac{H}{1+\frac{e^{6}}{3\pi}H}\right)=\lambda+2\ln\lambda+2\psi\left(1-\frac{1}{\lambda}\right)
  +\ln 2+4\gamma +\psi\left(1+|m|\right),
\end{eqnarray}
where $H\equiv B/(m_{e}^{2}e^{3})$ is the magnetic field in units
of atomic magnetic field, $\psi$ is the logarithmic derivation of the
gamma function and $\gamma=0.5772\dots$ is the Euler's constant.

For each $m=0,-1,-2,\dots$ the solution of (\ref{eq:405}) produces
a tower of states; the tower with the lowest values of $E_{\lambda}$
corresponds to $m=0$. The values of $E_{\lambda}$ in the limit
$B\gg 3\pi m_{e}^{2}/e^{3}$ (or $H\gg 3\pi/e^{6}$) are shown in
\cite{2}, Fig. 10.
However the atomic energies of the states from different towers are shifted by
the value $\Delta E_{m}=eB|m|/m_{p}$ and since $\Delta E_{m}$ grows
linearly with the magnetic field, for strong enough $B$ this shift is big.

Let us consider $B=2\cdot 10^{3}m_{e}^{2}e^{3}\approx 4.7\cdot
10^{12}$ Gauss which was used in the calculation of the energies in the
Table 1 of \cite{8}. For $m=0$ for the ground level from
(\ref{eq:405}) we obtain $\lambda_{0}^{0}=4.3$, $E_{\lambda}(m=0)=-255~
eV$. For $m=-1$ the ground level according to (\ref{eq:405})
corresponds to $\lambda_{0}^{-1}=3.8$, $E_{\lambda}(m=-1)=-193~ eV$
which is $62~ eV$ above $E_{\lambda}(m=0)$. The proton motion increase
the atomic energies of the $m=-1$ tower by $eB/m_{p}=30~ eV$ and the difference
of energies with account of the finite proton mass becomes $92~ eV$
(according to Table 1 of \cite{8} it equals $93.3~ eV$). For stronger
$B$ the second term in (\ref{eq:404}) starts to dominate over the
first one.

\section{Spin-spin interaction in hydrogen, heavy ions and positronium
in a strong external magnetic field}
\label{sec:hyperfine}

The interaction between the proton and electron spins leads to the
hyperfine splitting of the hydrogen atomic levels. Being proportional
to $\mu_{e}\mu_{p}|\psi(0)|^{2}\sim m_{e}\alpha^{4}(m_{e}/m_{p})$ it
is much smaller than the atomic energies, which are of the order of
$\alpha^{2}m_{e}$. In the case of a strong external magnetic field $B\gg
m_{e}^{2}e^{3}$ the spin-spin interaction considerably grows:
$|\psi(0)|_{B}^{2}\sim
|\psi(0)|^{2}\left(\frac{a_{B}}{a_{H}}\right)^{2}=|\psi(0)|^{2}\left(\frac{B}{m_{e}^{2}e^{3}}\right)$,
leading to a linear increase of the hyperfine splitting with
magnetic field\footnote{M. A. Andreichikov, B. O. Kerbikov, private
  communication}:
\begin{eqnarray}
  \label{eq:501}
  E_{SS}\sim m_{e}\alpha^{2}\frac{m_{e}}{m_{p}}\frac{B}{B_{0}}.
\end{eqnarray}

It follows that at $B\sim 10^{5}B_{0}$ the spin-spin interaction energy
becomes of the order of the freezing energy of the hydrogen ground level
$E_{0}\approx -1.7 keV$, and if a linear growth of $E_{SS}$ with $B$
would take place for $B>10^{5}B_{0}$ it would determine the value of
the ground state atomic energy $E_{0}$. However just at $B\sim
10^{5}B_{0}$ the Landau radius $a_{H}$ approaches the proton charge
radius and a power formfactor suppression of $E_{SS}$ occurs preventing
it from growing further. 

The energy of the spin-spin interaction in heavy hydrogenlike ions is
enhanced by a factor $Z$ which originates from the
$1/a_{B}=Zm_{e}\alpha$ factor in the expression for
$|\psi(0)|^{2}$. Since protons and neutrons from completely filled
nuclei shells do not contribute to the magnetic moment of the nucleus,
for $B\lesssim 10^{5}B_{0}$ the extra term in the energy is
considerably smaller than the value of the electron mass and the
consideration of nuclei criticality in strong $B$ does not change
substantially. (One should also take into account that the formfactor
suppression in heavy ions starts at smaller $B$).

It would be very interesting to understand to which shift of the
energy of the positronium ground state in a superstrong magnetic field 
the spin-spin interaction of the electron and positron leads. In the
absence of external magnetic field, due to this interaction, the
ground state of the parapositronium is lighter than the ground state of
the ortopositronium
\cite{11,12}:
\begin{eqnarray}
  \label{eq:502}
  E(^{3}S_{1})-E(^{1}S_{0})=\frac{7}{12}\alpha^{2}m_{e}e^{4}=
  \frac{7}{12}e^{8}m_{e}.
\end{eqnarray}

The behaviour of the positronium energy levels in an external magnetic
field has several specific features \cite{12}, and in fields $B\gtrsim
e^{8}m_{e}^{2}/e$ the state with lower energy is a mixture of
ortopositronium and parapositronium ground states in which the electron
spin is oriented in the direction opposite to the magnetic field, while
the spin of the positron is directed along $B$. Its energy shift due to
the spin-spin interaction is of the order of:
\begin{eqnarray}
  \label{eq:503}
  \Delta E\sim \mu_{e}^{2}|\psi(0)|^{2}\sim m_{e}e^{4}\frac{B}{B_{0}},
\end{eqnarray}
and for $B>10^{4}B_{0}$ the ground state positronium energy could
become lower than $-2m_{e}$ which should lead to
the production of $e^{+}e^{-}$ pairs from the vacuum
(see \cite{12a} as well).

However the spin-spin interaction Hamiltonian is determined as
a nonrelativistic expansion of the $e^{+}e^{-}$ scattering amplitude,
the expansion parameter being $p^{2}/m_{e}^{2}$ \cite{11}. In the case
of a strong external magnetic field $p^{2}/m_{e}^{2}\sim
1/(a_{H}^{2}m_{e}^{2})=B/B_{0}$, and for $B\gtrsim B_{0}$ the
correctness of the formulae (\ref{eq:501}) and (\ref{eq:503}) is doubtful.

Let us remind that the anomalous electron magnetic moment leads to a
linear growing with $B$ of correction to the lowest Landau level
energy. However this behaviour is valid only for $B\lesssim
B_{0}$, while for $B\gtrsim B_{0}$ the strong linear dependence on $B$ is
replaced by a weak double logarithmic one (\cite{16}, see also \cite{2}).

Concluding this section let us state that the behaviour of the
spin-spin interaction in atoms and positronium at strong external
magnetic fields $B\gtrsim B_{0}$ deserves further study.

\section{Conclusions}
\label{sec:conclusions}

In a magnetic field $B\gg m_{e}^{2}/e^{3}$ the potential of a
pointlike charge becomes screened due to large radiative
corrections. The screened potential has a Coulomb behaviour along the
$z$ axis at distances $a_{H}<|z|<1/\sqrt{e^{3}B}$ and $|z|\gtrsim
1/m_{e}$ and these regions define the ground state atomic
energy. Distributing the pointlike charge within a domain of the  
size of the nuclear radius $R$ leads to a less singular behaviour of
the potential at distances $|z|<R$. When $e^{3}B$ approaches $1/R^{2}$
the ground state energy approaches the limiting value. For hydrogen
the limiting value of $\lambda^{\rm gr}$ which defines the ground
state energy $E_{0}=-(m_{e}e^{4}/2)(\lambda^{\rm gr})^{2}$ is equal to $6.9$ 
instead of $11.2$ obtained in \cite{2} for a pointlike charge, and
$E_{0}^{\rm lim}=-0.65~keV$ instead of $-1.7~keV$.

The same phenomenon of going up of the ground energy level
is obtained numerically in the relativistic domain for hydrogen-like
ions. It leads to a nontrivial dependence of the critical 
nucleus charge on the magnetic field: the nuclei with $Z<60$ never
become critical while the ions with $60\leq Z< 210$ are
critical only within a finite range of magnetic fields. At $Z=210$ a
``second'' freezing energy reaches the lower continuum and the 
nuclei with $Z\geq 210$ are critical at any $B$.

\acknowledgements

We are grateful to M. A. Andreichikov, O. V. Kancheli, B. O. Kerbikov,
V.A. Novikov, and Yu. A. Simonov for valuable remarks and
discussions. This work is partially supported by the RFBR under the
Grants No. 11-02-00441, 12-02-00193 and by the Russian Federation
Government under Grants No. 11.G34.31.0047, NSh-3172.2012.2.

\end{document}